A Multi-GHz Analog Transient Recorder Integrated Circuit

By

Stuart Armin Kleinfelder

B.S. (State University of New York at Stony Brook) 1984

THESIS

Submitted in partial satisfaction of the requirements for the degree of

MASTER OF SCIENCE

in

ENGINEERING

ELECTRICAL ENGINEERING AND COMPUTER SCIENCES

in the

GRADUATE DIVISION

of the

UNIVERSITY OF CALIFORNIA at BERKELEY

Approved:

Chair: ...... *Robert G Meyer* ............ 5/11/92 ......
.................................................... Date
.................................................... 5/17/92 ......
.................................................... 5/19/92 ......

*****************************

A Multi-Gigahertz Analog Transient Recorder

Integrated Circuit

Copyright © 1992

by

Stuart Armin Kleinfelder

A Multi-Gigahertz Analog Transient Recorder

Integrated Circuit

by

Stuart Armin Kleinfelder

Abstract

A monolithic multi-channel analog transient recorder, implemented using switched capacitor sample-and-hold circuits and a high-speed analogically-adjustable delay-line-based write clock, has been designed, fabricated and tested. The 2.1 by 6.9 mm layout, in 1.2 micron CMOS, includes over 31,000 transistors and 2048 double polysilicon capacitors. The circuit contains four parallel channels, each with a 512 deep switched-capacitor sample-and-hold system. A 512 deep edge sensitive tapped active delay line uses look-ahead and 16 way interleaving to develop the 512 sample and hold clocks, each as little as 3.2 ns wide and 200 ps apart. Measurements of the device have demonstrated 5 GHz maximum sample rate, at least 350 MHz bandwidth, an extrapolated rms aperture uncertainty per sample of 0.7 ps, and a signal to rms noise ratio of 2000:1.

Dedicated to the memory of

the 25 Lost Souls

of the October 20, 1991 firestorm,

Oakland, California.

# Table of Contents

# List of Figures.

The kind support of Professors

Robert G. Meyer, Paul R. Gray,

Jan Rabaey, T. Van Duzer,

and Robert P. Ely, Jr.,

is gratefully acknowledged.

# 1. Introduction to Transient Recorders.

Transient recorders capture brief, high-speed analog transient signals for digitization and subsequent digital processing. Transient recorders are "single shot" devices - they acquire a transient in real time and do not require a repetitively occurring signal. Transient recorders can be distinguished from general analog-to-digital converters in that transient recorders usually cannot sustain continuous conversion. They may be limited to some maximum time over which uninterrupted acquisition can take place or to a maximum number of uninterrupted conversions. Transient recorders also require a trigger that initiates or stops acquisition. Without a trigger, the limited endurance of the recorder would not allow capture of most transients. Digitizing oscilloscopes are the most general example of transient recorders, though transient recorders are often designed to work in highly specific applications.

Two basic trigger schemes are employed by transient recorders, with many systems offering both. "Common start" trigger mode is employed when the timing of an analog transient can be predicted. This most often happens if the signal is regularly occurring. Using a common-start model, a trigger signal is applied to the transient recorder prior to the predicted arrival of the transient. With as brief a delay as possible, the recorder begins acquiring the signal of interest and continues until it runs out of fast storage. Afterward, the recorded data is examined to see if the anticipated signal actually occurred. This mode is called common-start because often multiple channels are triggered in common.

Common stop mode is employed when the arrival of the transient cannot be predicted. In this case, the recorder is started and runs continually, constantly looping over or otherwise discarding the oldest data while recording new data. A trigger circuit (i.e., comparator) eventually senses a transient, and the recorder is stopped. A delay is usually deliberately inserted between the trigger and the stop, so that the transient can be captured completely before recording is stopped (this is called post-trigger sampling). Similarly, pre-trigger samples should be kept because the trigger typically cannot respond with sufficient speed or sensitivity to start the recorder before the leading edge of the transient is lost. The experimental chip described here uses the common-start mode of triggering. While the common-stop mode has some added value, the overhead due to trigger capabilities are mostly irrelevant to the speed and analog characteristics of the system, the primary aspects of interest here.

Some of the principle criteria for effective transient capture are: maximum sampling speed, bandwidth, dynamic range, linearity, timing accuracy, record length, and minimum conversion time, dead time, power consumption and cost. Sampling speed is the most prominently quoted specification of transient recorders, it being the maximum sustained rate at which sampling can take place. Bandwidth is a measure of the responsiveness of the device to rapid changes of the analog input signal. Typically, bandwidth is quoted as the frequency at which a full-scale sine wave in the digitized output is found to be attenuated by 3 dB.

Linearity is quoted as differential and integral. Integral non-linearity is measured by determining the worst case deviation from a fit to the endpoints of the digitized transfer function. Differential non-linearity is a measure of the worst case error between any two adjacent samples in the transfer function. Minimizing aperture uncertainty or "jitter," noise in the timing of sampling or conversion, is important in that it directly influences the accuracy of conversion of high bandwidth signals. The record length specifies the maximum number of samples that can be acquired without interruption. Large record lengths are preferred, but many applications need a record length of only a few hundred samples. Conversion time is the time spent reading out or processing data. The conversion time is often the same as the so-called "dead time" of a system - time between sets of acquisitions during which the system is incapable of acquiring a new transient due to conversion, etc. Power consumption, space and cost are obvious and sometimes critical considerations, particularly in many instrumentation applications, where large parallel arrays of digitizers are needed. The characteristics of an oscilloscope capable of performing most tasks required by today's electronics community include at least 2 channel capability, at least 6 bits of displayed vertical resolution (eight preferred), at least 512 samples displayed, at least 250 MHz bandwidth, and at least 500 MHz sample rate. Exceeding these performance specifications is still quite difficult, though progress is presently rapid.

## 2. Methods of Transient Recording.

### 2.1. *Flash analog-to-digital conversion*.

Many techniques are employed in transient recorder systems. Flash analog-to-digital converters (FADC's) are commonly used today. They have the advantage of high sustained throughput, limited only by the depth of the memory storage system. Today, speed in practical systems has been limited to about 250 MHz without interleaving and 1 GHz with interleaving. Bandwidth has been limited to about 500 MHz. Vertical resolution is usually 8 bits in contemporary FADC based systems. Most modern FADC based digital oscilloscopes offer several thousand samples record length. Density is low, with 1 to 8 channels achieved per modular system (i.e., VMX board). Conversion time is fast, as the data is digitized on the fly and is essentially ready for processing immediately. Power consumption and cost are the highest of any alternative. Flash analog-to-digital conversion chips are presently evolving rapidly. Several experimental chips have been reported to operate in the few GHz range, albeit at lower resolution [1].

### 2.2. *Charged-coupled-device systems*.

Another fairly common technique, where lower speeds and depth are acceptable, is the use of charge-coupled-device (CCD) delay lines. These have been limited to about 50 MHz acquisition rate, a few hundred samples' endurance, and 50 MHz bandwidth, with vertical resolution as high as 10 bits. CCD's are difficult to clock, but have the unique advantage of having very low input capacitance, and therefore can potentially obtain high bandwidth. CCD's are very susceptible to temperature, clock rates, and other operating conditions, and require a fair amount of overhead and calibration. It is possible to imagine that, in high density applications, CCD based systems can cost well under $1,000 per channel.

*2.3. Synchronous switched-capacitor waveform samplers.*

Switched-capacitor technology has recently been exploited in transient recorder systems. An example is the EOS ("Equation Of State") Nuclear Science particle detector at Lawrence Berkeley Laboratory. This detector has 15,000 parallel transient recorder channels used to record energetic nuclear collisions. The 15,000 channels use the SCA ("Switched Capacitor Array") transient recorder circuit designed by this author [2,3,4]. Each SCA chip contains 16 parallel channels, each able to record 256 samples at up to 100 MHz. The maximum acquisition speed of such systems is comparable to CCD's and non-interleaved FADC's, the memory depth is comparable to CCD's, and the bandwidth is comparable to FADC's. The dynamic range, at 8000:1, is superior to either CCD's or FADC's. Cost, in a large system such as EOS, is about $100 per complete channel - perhaps less than any other similar type of system. The cost of the SCA alone was only $4 per channel. Power consumption was only 10 mW per channel. Note that dynamic range is quoted, rather than bits of resolution, because the circuit does not actually digitize the analog information. Rather, an external converter is used, at whatever resolution is desired, to digitize the reconstructed analog signal. Because of this, the recorder itself does not suffer from quantization error.

Switched-capacitor-based systems can permit much higher flexibility and creative operation when compared with a simple linear CCD. For example, a subsequent SCA device, also designed by this author [5,6,7], has the features of dual ported random access for acquisition and readout. This permits all the flexibility in addressing of a conventional digital memory. Performance figures for this device, developed for use at the Superconducting Super Collider, are listed in table 1.

Table 1. Performance of the SSC Switched Capacitor Array I.C.

| | |
|---|---|
| Acquisition speed: | > 100 MHz |
| Acquisition bandwidth: | ~ 100 MHz |
| Noise level: | 0.5 mV rms |
| Cell to Cell variation (noise averaged out): | 0.7 mV rms |
| Integral non-linearity: | 0.2 % |
| Supply voltage: | 5 V |
| Input range: | 0 V - 5 V |
| Output range: | 0.2 V - 4.8 V |
| Dynamic range: | 76 dB rms. |
| Power consumption per channel: | 10 mW |
| Number of parallel channels per chip: | 16 |
| Number of sample and holds per channel: | 256 |
| Acquisition mode: | Addressed |
| Readout mode: | Addressed |
| Readout speed: | 500 KHz |
| Technology: | 1.2 um CMOS |
| Die size: | 4.5 x 6.5 cm |

It is certain that SCA type circuits can be improved over time. Two-way interleaving has been used but not pressed to its limits. It is likely that 250 to 500

MHz synchronous sampling can be attained. Four-way interleaving may allow one GHz sampling. Bandwidth can likely be increased to about 250 MHz without serious difficulty. A dynamic range of 12 bits has been obtained without the need for corrections. This might be increased by a number of subtle improvements or with the use of a fully differential analog signal pathway. Endurance of 512 samples has been achieved, but this can be increased to 1024 or 2048 samples per trigger with some trade-offs. Density (silicon area used per sample) is likely to improve slowly. On-chip analog-to-digital conversion and digital buffering are being incorporated. This technology is an active area of research in the physics community due to its flexibility, high density, low cost and low power consumption. The purpose of the research described below is to probe some of the extremes of performance possible using the switched-capacitor waveform recording technique.

## 3. Fast Switched-Capacitor-Based Transient Recording.

The switched-capacitor sample-and-hold technique used for the gigahertz circuit described below is a spin-off from work on the high energy and nuclear science instrumentation developments mentioned above. A diagram of the system architecture is shown in figure 1. Five hundred and twelve sample and hold NMOS switch and capacitor combinations are arranged in a dense linear array. The individual sample and hold cells consist of two transistors and a high density double polysilicon capacitor. The capacitor is approximately 952 square microns in area. With 60 fF per square micron, the capacitor is calculated to have a capacitance of 0.57 pF, ignoring edge fringing. Both switches are NMOS with a 1.2 micron drawn gate length. One switch is used to read out the cell, and is of minimum width, 2.4 microns. This is to reduce the undesirable voltage sensitive parasitic capacitance on the sample-and-hold node. The analog sample (write) switch is a wide device, 38.4 microns as drawn. It is folded (constructed using 2 physical gates to yield one logical gate) to reduce drain capacitance. The source diffusions are also shared by adjacent cells to reduce source capacitance.

The write clock is generated by an interleaved, asynchronous delay line system, described below in some detail (see figure 2). A single switch could not be used to both write to and read from the capacitor, as the overlapping, asynchronous write technique used is incompatible with the much slower synchronous read technique used. NMOS sample and hold switches alone are used instead of CMOS transmission gates, due to the extreme speed of the part and the difficulty, at such speeds, in matching the timing of complementary clocks. As the minimum pitch of the sample-and-hold cells is fixed by the clock wiring pitch, using CMOS switches, with twice the number of clocks per sample and hold cell, would have halved the pitch of the sample and hold cells. The use of NMOS switches instead of CMOS switches causes about a 40% penalty in signal range due to the substantial body effect of the NMOS transistors when a large source voltage is imposed.

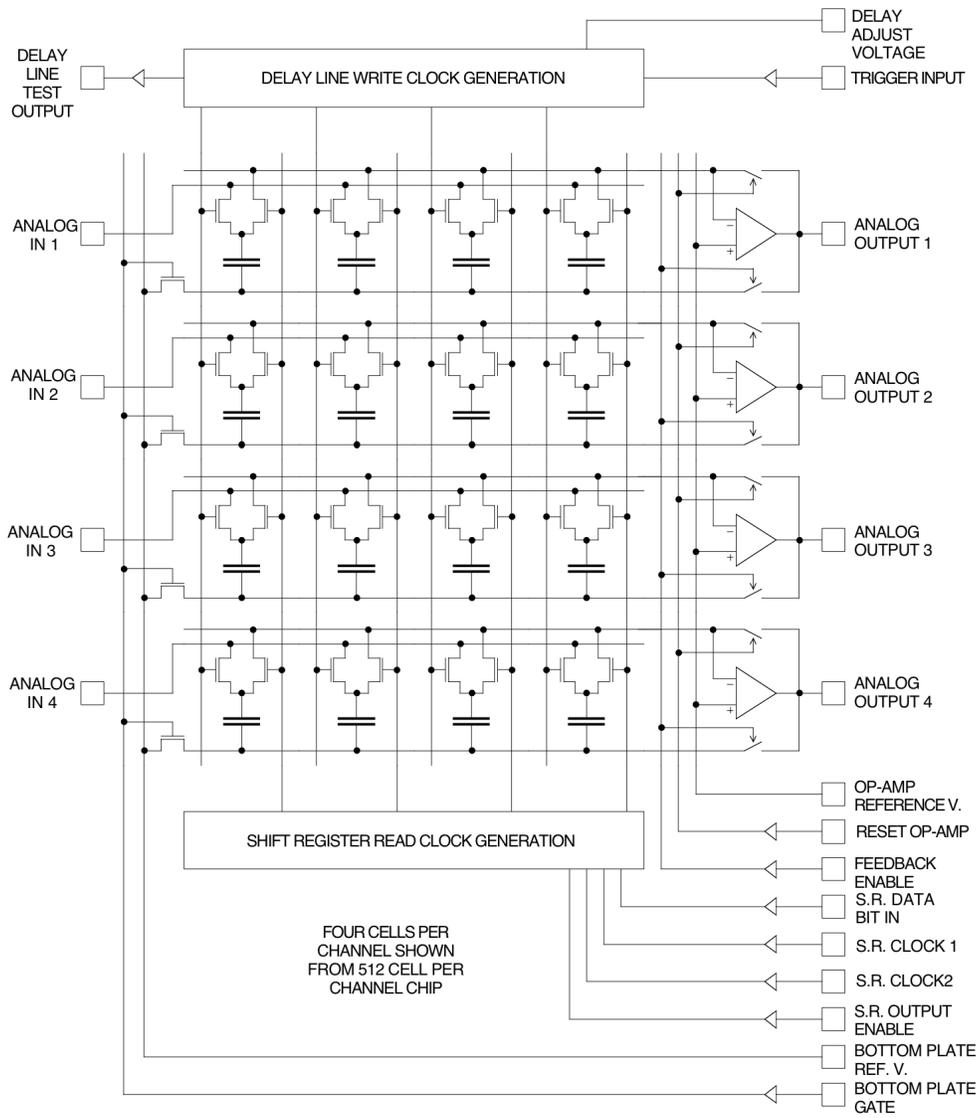

Figure 1: Transient recorder architecture.

The bottom plates of all capacitors within a channel are shared, and shorted together permanently. Two switches are used to multiplex the purpose of the bottom plate bus. A quite large (614 over 1.2 micron) NMOS switch applies a reference voltage to the bottom plates during acquisition, and a small (12 over 1.2 micron) NMOS switch connects the bottom plate to the output of an operational amplifier during readout. This op-amp is a multi-stage rail to rail amplifier design after reference [8], and proven to operate adequately on a single 5 Volt supply in the 1.2 micron technology used.

Two buses per array of sample and holds distribute the analog input and output to all cells. The analog output bus connects all sample and holds by their read transistors to the readout op-amp inverting input. A two phase dynamic CMOS shift register based serial to parallel converter is used to clock the S&H read switches during readout. A break-before-make action is enforced by using a third clock to gate the parallel outputs of the shift register. The shift register circuit lies dormant during acquisition.

The sample reconstruction technique shown in figure 3 switches' sample capacitors, one at a time, into the feedback path of the readout amplifier, configured as a unity-gain inverting follower.  This is a technique similar to that discussed in reference [9].   A reset switch can short the op-amp inputs and outputs, balancing the op-amp and removing parasitic charge remaining on the bus between the readout of each cell.  An inverting configuration is used to permit the  low impedance amplifier output to be switched to the shared bottom capacitor plate.  With its high capacitance to the die substrate, connecting the shared bottom plate to the high impedance op-amp input would be an invitation to high levels of pickup and noise.

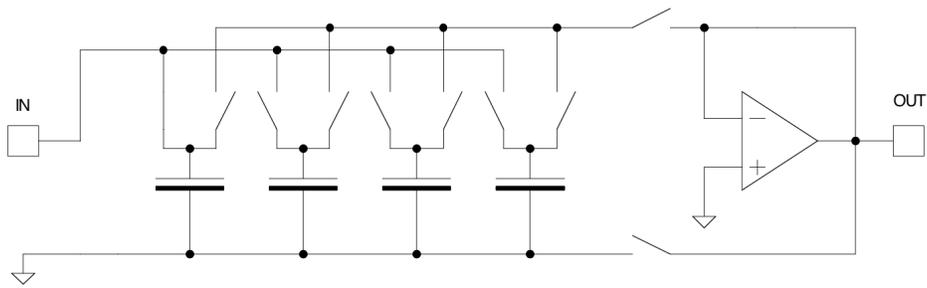

(A) Start acquisition of first sample.

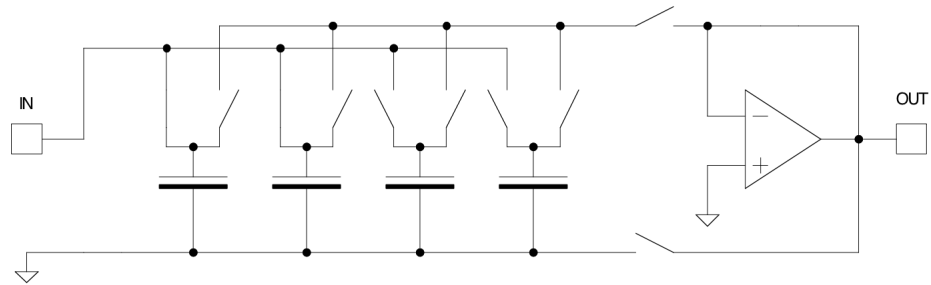

(B) Start acquisition of second sample.

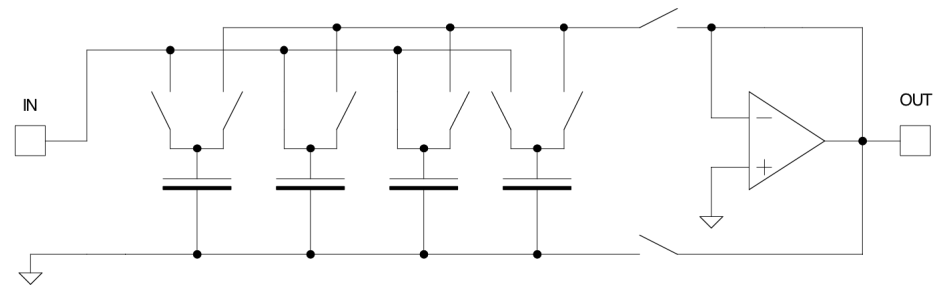

(C) Complete first sample, start third sample.

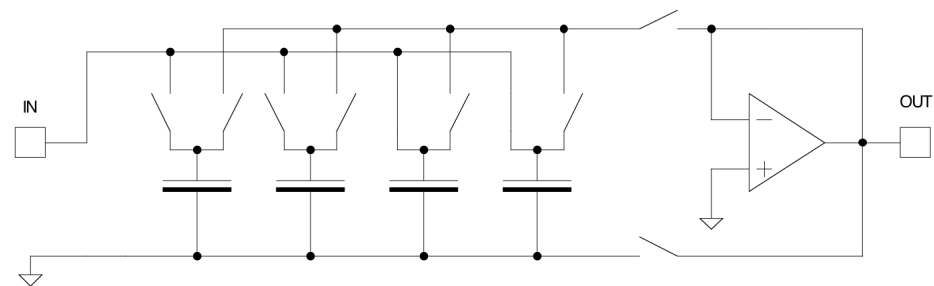

(D) Complete second sample, start fourth sample..

Figure 2: Write cycle using 2-way interleaving.

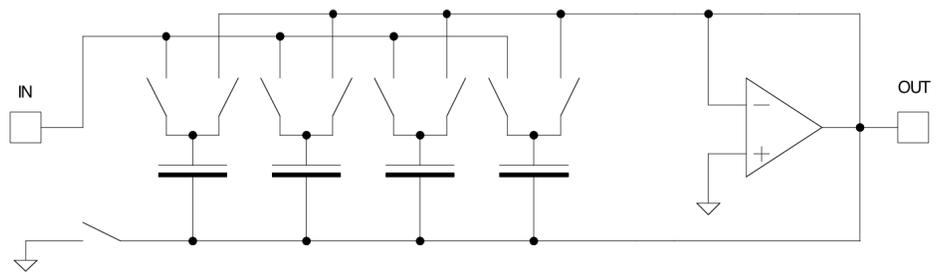

(A) Auto-zero readout path.

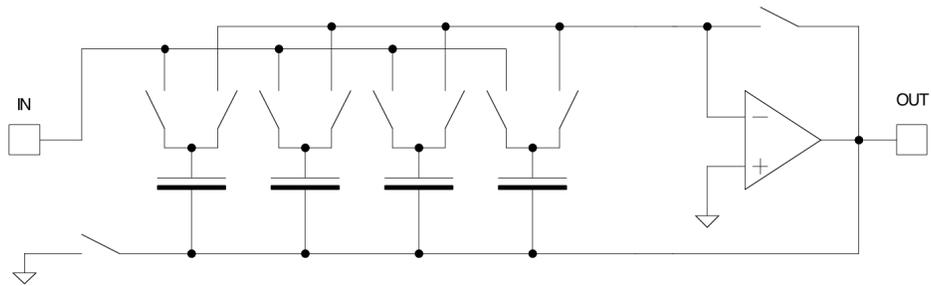

(B) Break reset, ready to read first cell.

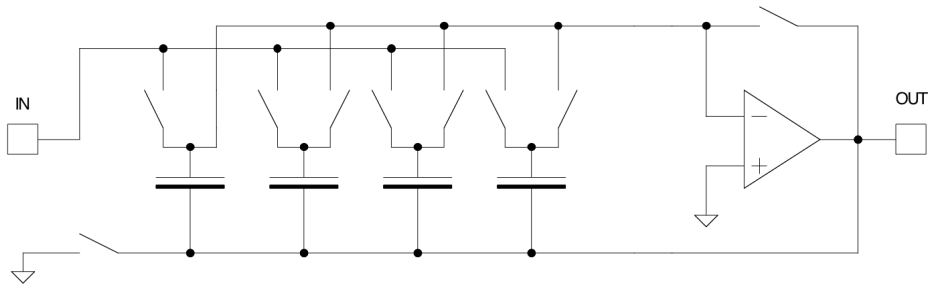

(C) Read first cell.

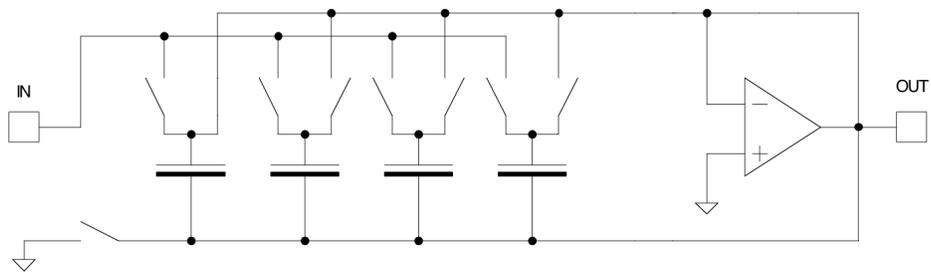

(D) Auto-zero capacitor and bus parasitic.

Figure 3: Read cycle.

## 4. Bandwidth Tradeoffs.

The analog bandwidth of the circuit depends on several factors. Most obvious is the RC time constant of the sample and hold switch and capacitor. The channel resistance is strongly related to the gate to source voltage, and therefore the analog input voltage:

$$R_{ch} \approx \frac{1}{\mu C_{ox} \frac{W}{L} \left( V_{gs} - V_T \right)}$$

A summary plot showing the results of calculations of bandwidth verses input voltage for the topology used in the fast transient recorder is shown in figure 2. At low input voltage, where Vgs is highest, the simulated bandwidth (-3dB) is 1.8 GHz, a reasonable match to the 5 GHz sample rate achieved. Operation up to about 2.5 volts is seen to yield a bandwidth of 1 GHz or higher.

More insidious is the effect of the input capacitance of the chip in combination with the output impedance of test sources and the inductance of the device socket, package and bond wire. The on-chip input capacitance (no package, socket, etc.) was measured to be 15.5 pF. Given this capacitance, a 50 Ohm output impedance signal source would result in a 0.78 ns time constant and a 1.7 ns 10% to 90% rise time. This corresponds to only 200 MHz bandwidth - much worse than the sample and hold bandwidth. Reducing the output impedance of the test source can help with this, but adding in only a few nH of inductance also limits bandwidth, as well as causing predictable resonance effects. Bandwidth proved to be one of the most frustrating measurement challenges, in view of the hints of very high performance potential. Indeed, a very high performance part requires special low inductance packaging and highly optimized analog input buffering. In the end, a practical system may need to trade off sample and hold bandwidth to obtain a lower input capacitance, more user friendly system.

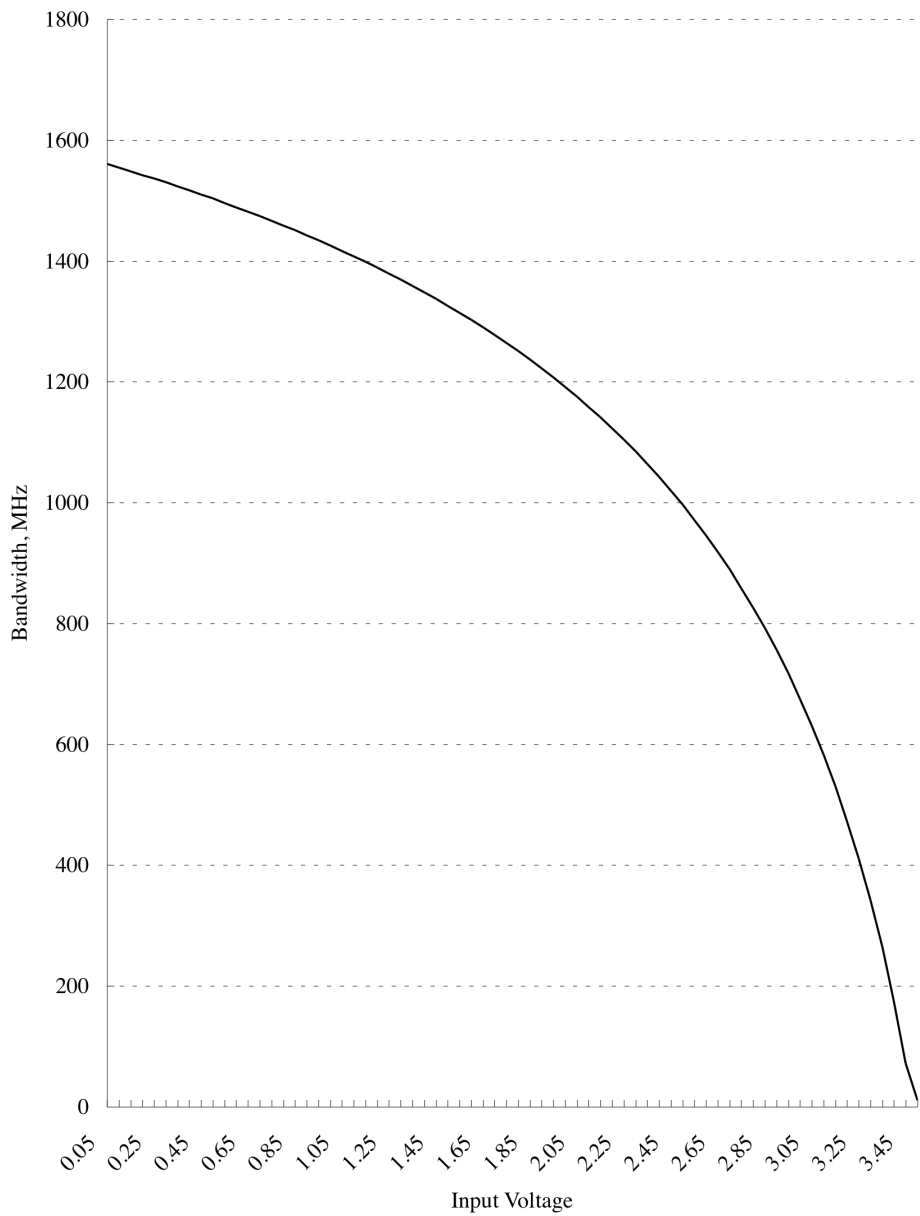

Figure 4: Sample and hold bandwidth vs. input voltage.

# 5. Achieving Fast Sample Rates via Asynchronous Clocking.

*5.1. Active delay lines.*

One of the interesting experimental techniques demonstrated in the test device is the delay line write clocking. Achieving multi-gigahertz speed given the limitations of a moderately advanced CMOS process is difficult. A clocked system at a sample rate of 1GHz or higher is almost impossible without the use of architectural expediencies such as interleaving. Even if four-way interleaving were used to reduce a 1 GHz clock down to four 250 MHz clocks, the system would remain difficult to design and use. The delay line clocking technique may be a workable alternative in many applications that avoids the high speed synchronous clocking altogether, while opening the possibility of extremely high acquisition rates.

For common-start use, an open ended string of tapped delay elements can be used as a basis for rapid clock generation. An example of such a circuit would be a long string of inverters. An edge or pulse injected into the start of the string propagates down the string until it reaches the end, where it is extinguished. An open ended string is only useful when the position of the transient can be predicted, for the trigger must precede the transient. This type of circuit is simple to operate because the relationship between time and space is fairly clear. By this it is meant that the first tap in the string will switch just after the trigger arrives, and the last tap will switch a predictable amount of time later. There is less opportunity for ambiguity as to which samples in time correspond to which physical locations in the string. Open ended strings in a common-start environment can allow easier pre-conditioning of the sampling electronics. For example, sample and holds can be pre-charged to a known value prior to triggering the acquisition. This can eliminate any memory effect - the influence of prior conditions on the recorded signal. Depending on the specifics of the circuit, an open-ended string may suffer from different behavior of the beginning, middle, or end most delay elements.

A ring shaped tapped delay line is more useful in a common-stop triggering environment. A ring can be caused to oscillate - the trigger stimulus launched and propagated in a circle indefinitely until deliberately caused to stop. As the ring oscillates, the input signal continually overwrites previously stored samples. After the transient is recorded, the stop trigger halts the oscillation. The oscillation must be halted quickly lest the acquired transient be overwritten and lost. The position of the sample at which the stop arrived must be carefully noted because the correspondence between space - the physical location of a given sample - and the time at which that sample was acquired is not otherwise clear

*5.2. Single-ended verses fully differential.*

Simple inverters have different rise and fall times, and different propagation times depending on the edge direction. Therefore, taps at every inverter output would produce quite poor tap-to-tap uniformity in delay. Taps based on inverter pairs would deliver high tap-to-tap uniformity and have the practical advantage of being non-inverting. Fully differential inverters would permit the greatest flexibility, but only if, in response to a differential input, the two outputs swing differentially with exactly 180 degrees of phase difference. Otherwise, the situation would be similar to the single-ended inverter string case. The most workable possibility would be the use of simple fully differential low-swing, low-gain amplifier stages. These differential amplifiers would consume power continually, whereas in a string of CMOS inverter pairs the power consumption would be zero for all inverters save those in transition. Spice simulations revealed the possibility that pairs of inverters could propagate signals within 200 ps (100 ps per inverter) - a figure hard to beat by the use of linear amplifiers. Therefore, the advantages of speed, simplicity, low power consumption and robust design verification lead to the use of delays per tap consisting of single ended inverter pairs.

*5.3. Single-ended options.*

Options for CMOS inverter pairs include simple inverters, ratioed CMOS inverters or dynamic inverters. Simple inverters can have approximately symmetric output characteristics. Ratioed inverters can be designed to favor propagation in one transition direction at the expense of the other. The same holds true for dynamic inverters. Dynamic inverters have drawbacks in a sensitive mixed analog and digital circuit. They can produce large transient power draw when being precharged and entering evaluation. Prior to entering the evaluation phase, they can display output offsets due to charge injection from the clock precharge gate. Also, there is the additional overhead of the clock that is not present in the static inverter case. Dynamic logic may be faster however, and this is what is most coveted.

Ratioed CMOS inverter pairs, where every other inverter is designed to favor one transition edge, can closely approach the speed of dynamic circuits. Dynamic circuits can be viewed as ratioed circuits taken to the extreme. In the transient recorder test circuit, heavily ratioed static logic was chosen. The ratios used resulted in approximately 50 times higher speed in one direction than the other, and essentially reached the speed potential (in one direction) of dynamic logic without any of their drawbacks.

*5.4. Control of tap delays.*

An issue arises as to how the speed of the delay line can be actively controlled. Lack of any control would leave the speed per tap at the mercy of the fabrication intrinsics', temperature variations, etc. An obvious possibility is an analogical control. This was achieved in the test circuit by adding a single additional transistor per tap, whose gate is externally controlled (see figure 5). The two inverters per tap were divided into a "slow," controlled fall time inverter and an uncontrolled fast rise time inverter. The fast inverter is a conventional CMOS inverter, but one whose transistor ratios are modified for fast rise times at the expense of slower fall times. The slow inverter is ratioed for fast fall times at the expense of slow rise times, but also contains an additional transistor between the n-channel source and ground. The gate of this transistor can modulate the current available to the inverter during the inverters high to low transition. By starving the inverter for current, the speed of the falling edge can be controlled. The fast rise time inverter, on the other hand, sharpens the slower edge of the first inverter. Ultimately, an on-chip delay stabilizing feedback system would be needed to make the speed of this chip insensitive to temperature, etc.

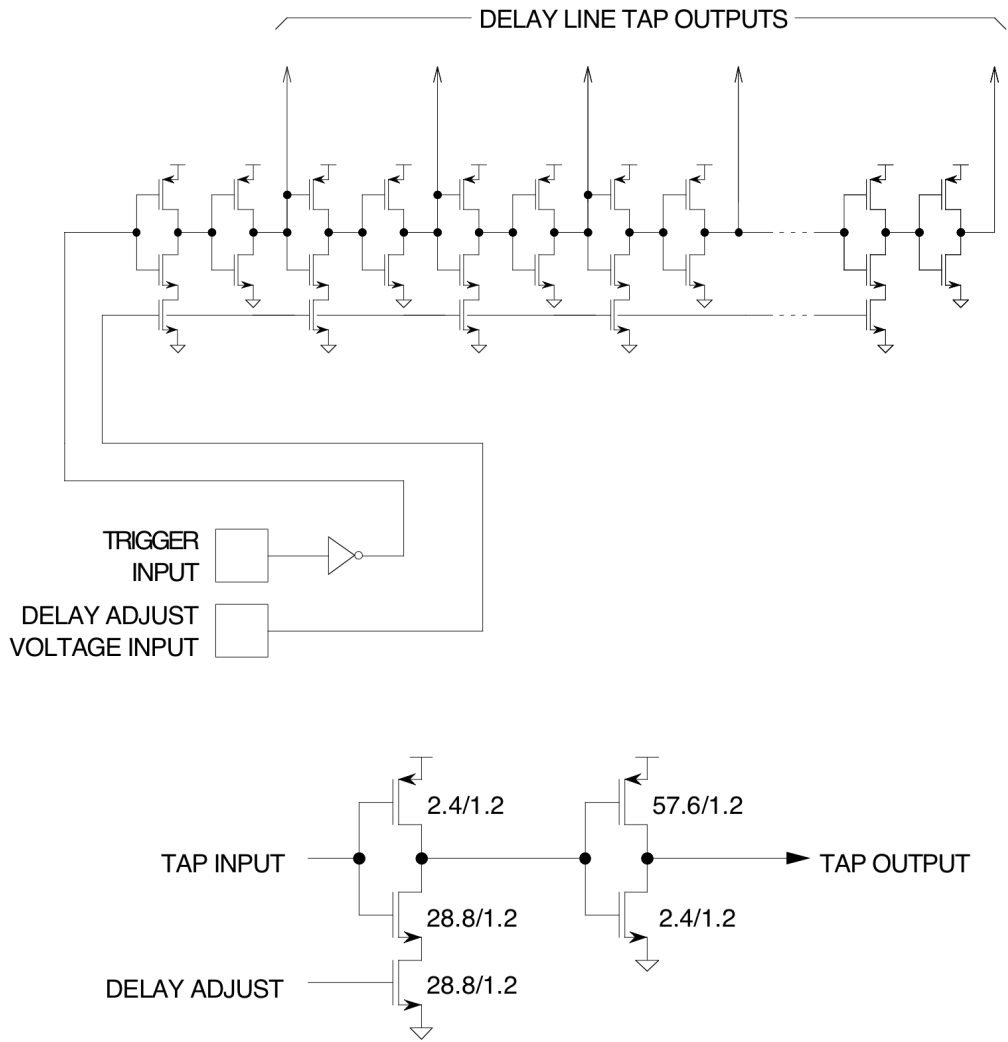

Figure 5: Voltage controlled delay line circuit diagram.

6. Active Delay Line Write Clocking Using Look-Ahead.

The delay line clocking technique uses a deep, analogically adjustable, tapped delay line to generate the hundreds of individual sample and hold clocks. In the experimental chip, 512 samples can be taken per trigger on 4 parallel channels, for a total of 2048 samples. A 512 tap deep delay line is used to generate the 512 clocks. In actuality, 528 taps are included, but the first 16 are used only to launch the acquisition with greater uniformity. Each tap in the delay line is composed of two inverters, the speed of one of which is analogically controllable. With the analog control, the delay per tap can be tuned to be 1 ns or whatever is desired within circuit operation limits.

A circuit that left all sample switches closed initially, and then opened them sequentially, would cause the input capacitance of the system to vary dramatically over time. In the test chip, it would initially equal the analog signal distribution bus parasitic plus 512 sample capacitors. At about 0.57 pF each, the input capacitance would start out to be over 290 pF - very large. At the last sample, the capacitance would equal the chips parasitic input capacitance plus only one sample capacitor, many times less. This disparity would likely result in poor analog behavior. Instead, a sample and hold scheme is used that requires each sample and hold switch to be closed briefly and then opened sequentially. This keeps the input capacitance lower and almost constant over time.

If, by using the delay line write clocking scheme, the achievable time per sample is only a few hundred ps, there is the severe problem of generating the clocks to close and open the hundreds of sample and hold switches. The problem is that the CMOS technology is hard pressed to swing from low to high and back to low in the few hundred ps available. To solve this problem, it was necessary to permit several sample and hold switches to be closed at the same time; a form of interleaving. The interleaving may adversely affect analog performance. Observing its effects is an important part of this investigation.

Propagating a sample and hold clock pulse down a digital delay line is possible, but there is little expectation that good uniformity could be achieved. For example, the matching of the propagation speed of the leading and trailing edges is likely to be poor even with both actively controlled. This would inevitably result in the widening or narrowing of the pulse as it propagates. In an extreme case, a faster trailing edge could catch up to the leading edge entirely, and the pulse would be extinguished. The test device avoids this by propagating an edge only. A look-ahead technique was devised that permitted a single edge to initiate both switch closure and opening (see figure 6). Following the delay line taps are a series of inverters and NOR gates that combine and buffer the tap clocks. The NOR gates are used to produce a look-ahead action that closes a sample and hold switch some period of time prior to the arrival of the clock edge that opens the switch. The final inverters are used to control edge speeds and buffer large clock load capacitance. All transistors in the delay line and buffering are ratioed to favor short delays during the write cycle and fast falling edge speed when opening the S&H switches. Achieving a short fall time on the S&H switches is important to the sampling accuracy of high swing, high bandwidth signals.

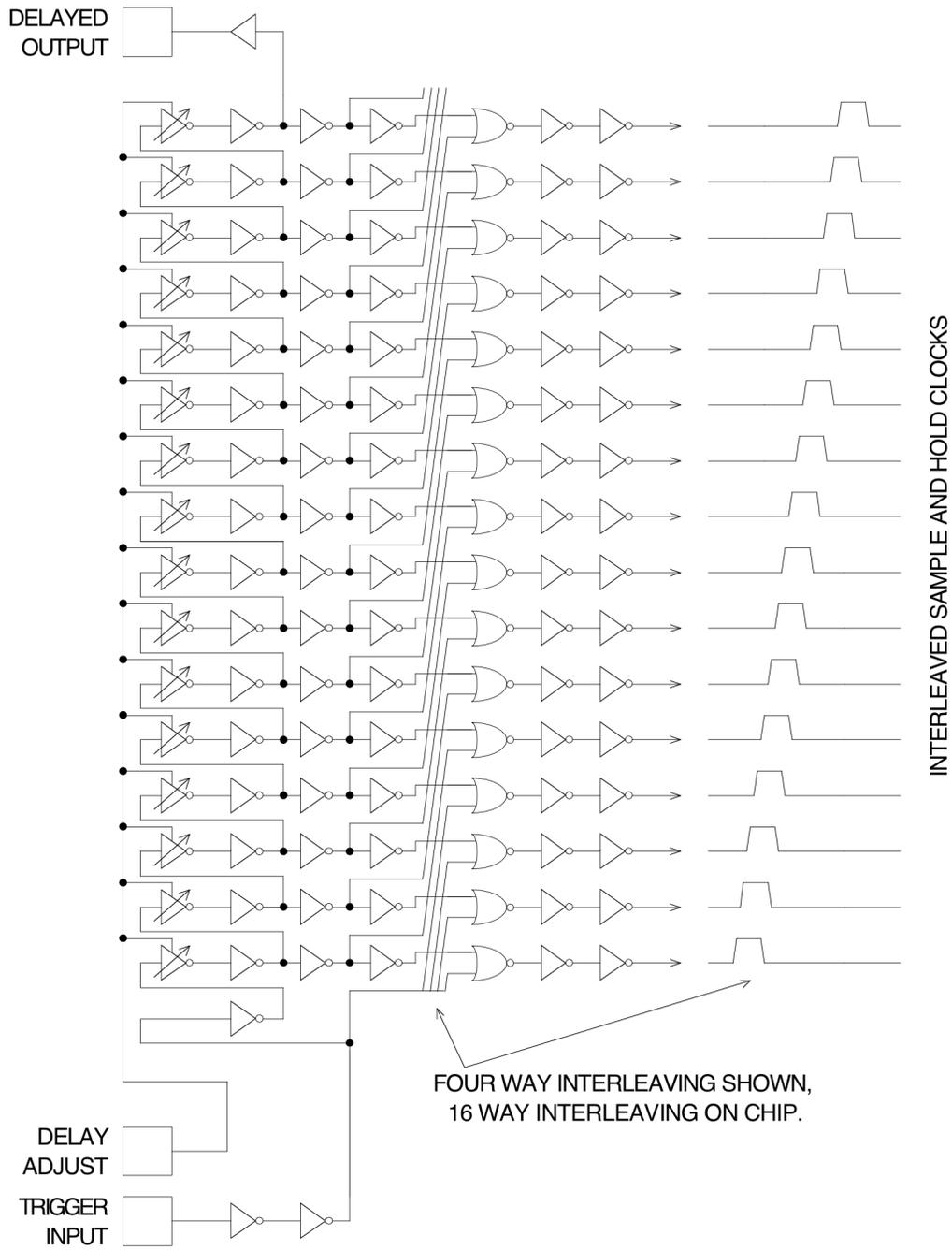

Figure 6: Interleaved Asynchronous Clock Generation.

# 7. Layout Issues.

## 7.1. *Layout issues for uniform timing.*

The 512 sample and hold cells are physically arranged in an unbroken, completely uniform, straight line to prevent systematic geometrical non-uniformity's from corrupting analog performance. It has been found in the past that any mirroring, rotation, or cell to cell non-uniformity are ultimately visible in the analog performance of the part. The pitch of the S&H cells is 13.2 long by 180 microns wide, with 512 cells occupying 5533 by 180 microns of area. A floor plan of the experimental chip is shown in figure 7, followed by a die photograph in figure 8.

At least the very first or very last taps may have different loading conditions, causing them to behave differently. It is not uncommon for transient recorders to acquire more samples than are kept due to discontinuities in the behavior of the first or last samples. A question arises as to what form of non-uniformity would be the least disconcerting to a user in a general application - i.e., an oscilloscope user. Psychology is as important here as are mathematical measures of accuracy. It is speculated that a layout that reduces the peak non-uniformity and gross patterns in non-uniformity (for example, a visible ramp, bow or saw-tooth pattern) at the expense of fine grained non-uniformity (i.e., odd-even effects) would be less noticeable and annoying to most users.

A Delay line in the form of an open ended string is best organized logically and physically as a straight line. A layout that has bends, corners or jogs will suffer slight non-uniformity in timing. Mirroring every other samples electronics would cause an odd-even effect to be visible in a measurement of timing uniformity. Corners would cause unusual timing uniformity for one or two samples. Bending the delay line into a U or S shape may cause a saw-tooth or triangular pattern of non-uniformity in timing. A ring shaped delay line obviously cannot be accomplished without forming the layout into a loop. A ring formed as a straight line of delay elements with a returning conductive path would certainly exhibit a severe discontinuity in timing uniformity at the point of return. If formed into a perfect circle, no geometrical variations would exist, but some variation in timing would still be expected due to process gradients.

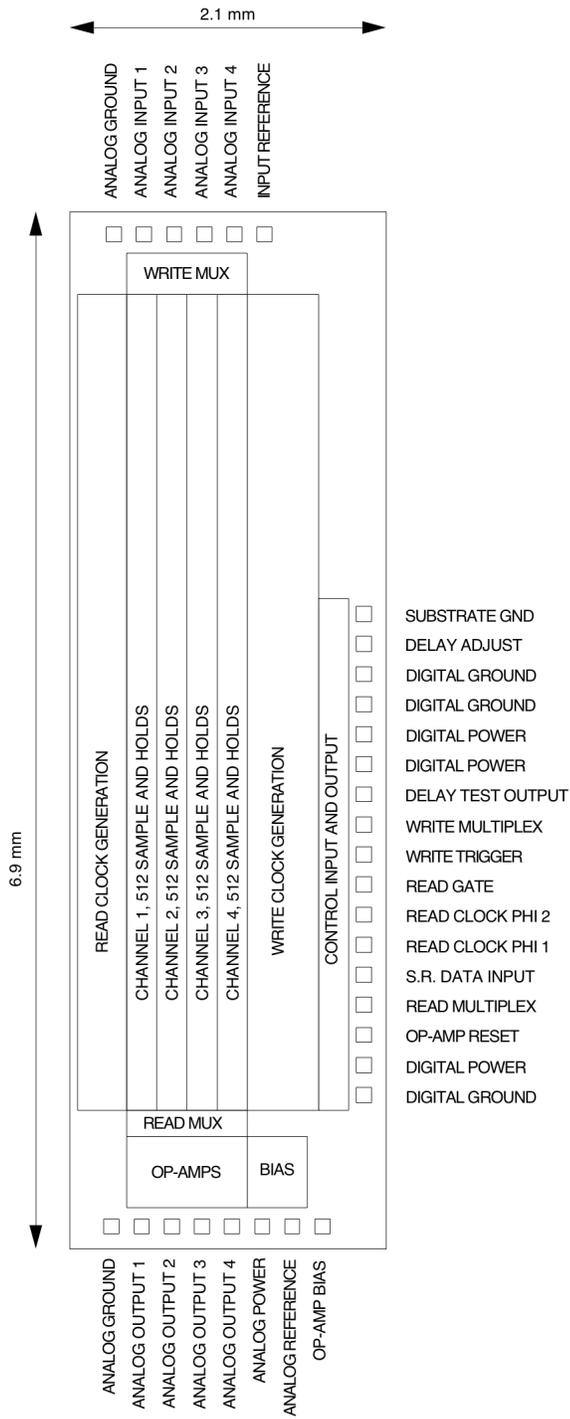

Figure 7: Transient recorder I.C. floor plan.

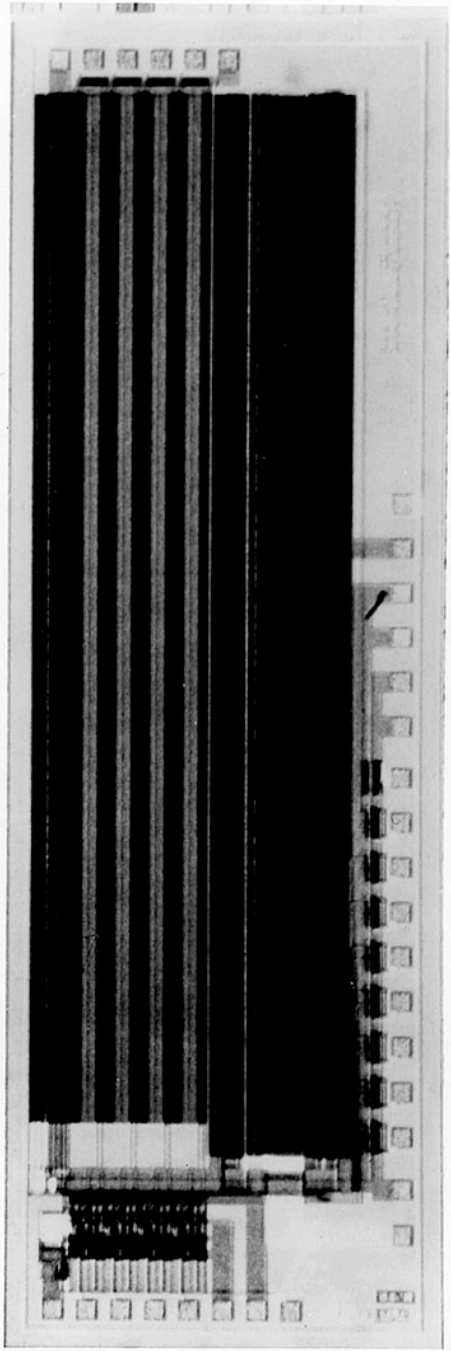

Figure 8: Plot of the fast transient recorder chip, containing 4 parallel channels of
512 samples each. The 1.2 micron CMOS die is 2.1 by 6.5 mm in size.

The fabricated test device used an unbroken string in which all sample and hold electronics used completely uniform geometry. Any non-uniformity would be limited to the first few samples, the final samples, or a gross pattern such as a ramp or bow across the 512 samples. Indeed, the most visible non-uniformity was seen at the final 16 samples in the string. The final 16 delay line elements were loaded with slightly less capacitance than others, as there were no succeeding elements in their look-ahead path. These last 16 elements had slightly faster response. The first 16 samples were not noticeably different due to the use of the extra "launcher" delay elements. Thus, an open ended string would benefit from the use of extra "lander" delay elements in addition to the successful launcher taps. Were a ring shaped system used, there would be no need for these extra taps as long as the first few samples after the start of oscillation are ignored.

*7.2. Geometrical uniformity in analog cells.*

The sample and hold cells, op-amps, etc., are at least as sensitive to uniformity issues as the clock logic. The effect of non-uniformity in the analog subsections may not as readily influence timing, but will affect cell to cell or channel to channel matching. For example, mirroring every other sample and hold cell could be expected to result in a visible odd-even effect in cell to cell baseline matching. In the high speed test circuit, all sample and hold cells were oriented in a uniform, unbroken line, without mirroring or rotation. As a result, no pattern in cell to cell baseline or gain could be found. Instead, any non-uniformity was confined to random variation and small semi-systematic effects that are most plausibly due to gradients in process chemistry or temperature across the die.

*7.3. Electrical uniformity.*

In addition to geometrical uniformity, electrical uniformity can be important. For example, the long metal lines that supply power may cause voltage drops across a row of delay elements. These voltage drops would cause the speed of taps to vary according to their distance from the supply pins. This would be most apparent in the case of delay line elements that use D.C. power, but may be detectable even in the case of zero D.C. power CMOS delay elements if the timing could be analyzed with sufficient accuracy. In the test chip, power busses were kept as wide as possible to reduce the chance of observing this effect.

# 8. Test Procedures and Results.

## 8.1. *Chip-on-board test jig*.

A small printed circuit board (6 cm by 6 cm), seen in figure 7, was designed and fabricated to facilitate tests of the experimental chip. For better performance, the "chip-on-board" technique was used. The die is mounted directly onto a small pad on the board. Wire bonds stretch a small distance from the chip to printed fingers on the board. To facilitate secure bonding, the board was gold plated. After mounting and bonding the die, a drop of fast setting urethane plastic was applied to encapsulate the die and bond wires. This may help protect the circuit from the environment, though the long term effect of chemicals in the encapsulant is not known. Apart from the chip-on-board technique used for the test device, conventional plated through holes were used for all other active and passive components. Other components on the board included TTL to CMOS Schmitt trigger buffers, screw-adjustable resistive divider bias voltage supplies and decoupling capacitors. Analog and digital ground planes were located on the bottom of the board. A dual row in-line flat cable header was used to allow attachment of off-board control logic. Analog signal inputs use small Lemo 50 Ohm coaxial cable connectors, with termination and optional voltage division. Analog outputs were available on simple test clip posts. The resulting board layout was satisfactory for many tests, but did not permit the full potential of the test device to show. A better board would use controlled impedance (i.e., strip-line) routing of the analog inputs, with small surface mounted termination resistors. Instead of wire bonding, ribbon bonding over as short a distance as possible, preferably starting from on top of the termination resistor, to the test device, would decrease lead inductance.

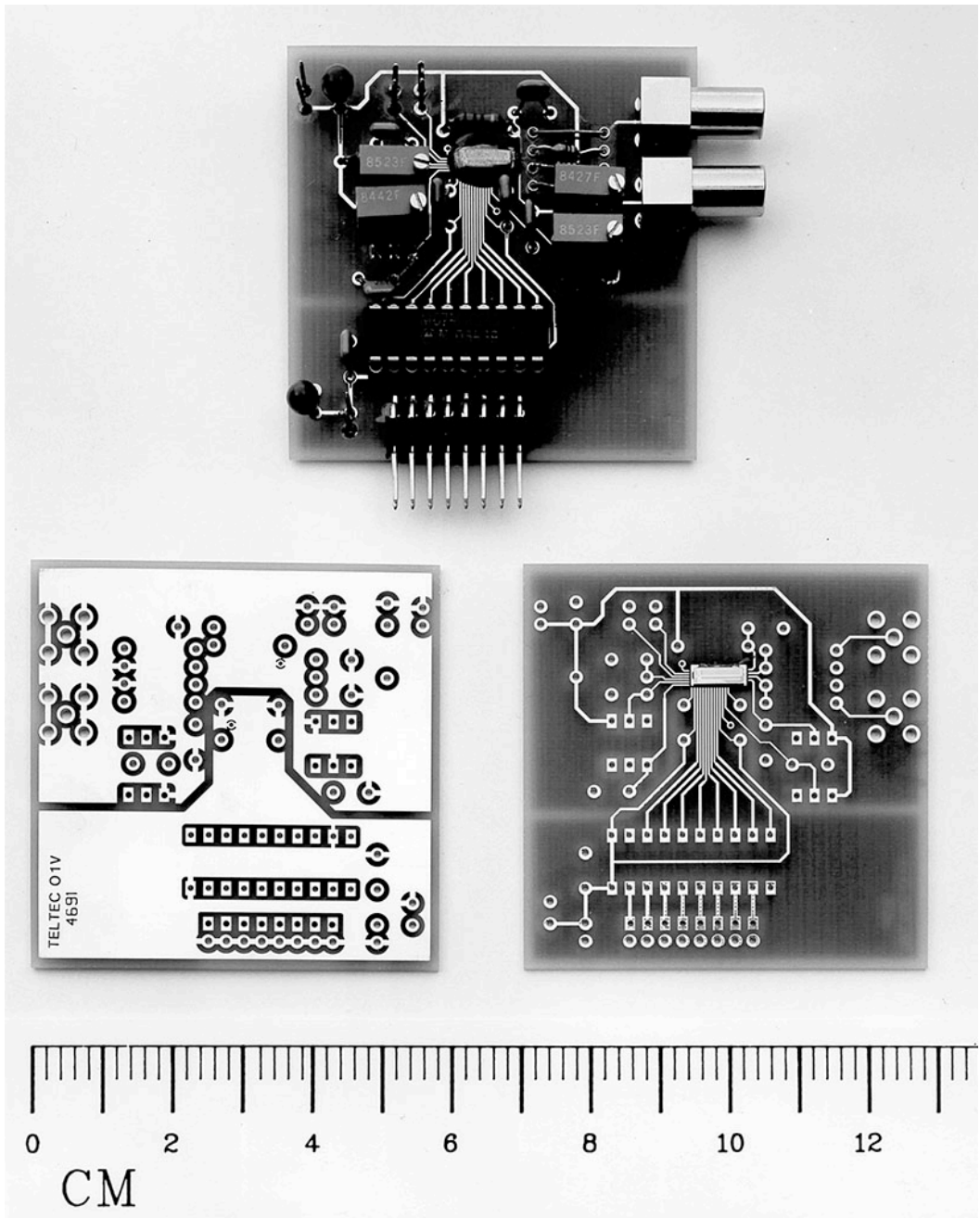

Figure 9: Chip-on-board test jig. The completed board is shown at the top, the left shows the bottom of the board and the right shows the top of the board with a chip mounted. The chip is the small rectangle at the upper center of the board.

*8.1. Maximum sample rate.*

The maximum sample rate was measured in several ways. The chip design included a test output - a done signal - that delivers an edge at the conclusion of the 512'th sample. An oscilloscope (Tektronix 4145) was used to measure the difference in time between the insertion of the common-start trigger edge, and the production of the done signal. This time span included the latency of a two stage input receiver, the 16 "launcher" tap delays, the 512 true sample delays, and a four stage output driver. If the launcher taps are considered equal to the true delay taps, then a total of 528 taps exist between the I/O ports. A separate input receiver and output driver pair, configured in a loop, gave a measure of their latency. The total latency, and a little math, result in an easy measure of the sample tap delay, and hence of the sample frequency. If there are small errors due to different launcher or I/O loading effects, the error (which cannot be more than a few nanoseconds' total), when divided among the 512 samples, remains small. An oscilloscope photo, showing an example measurement of this type, is seen in figure 10. The delay of 99.8 ns, divided by 528, demonstrates that a minimum of 5.29 GHz performance was reached.

A second measurement technique was to insert a wide pulse of well-measured width (from a 300MHz programmable pulser) into the sampler chip, and count the number of samples obtained in the read-out pulse. The width of the pulse divided by the number of samples in the recorded pulse yields a figure for the time per sample. Since it was not possible to obtain rise and fall times short enough to fall entirely into one sample, some judgment was necessary in determining where the pulse edges began. Therefore, there is a possibility of error in this measurement technique as well. The error is minimized by making the pulse as wide as possible, thus spreading any error over more samples and reducing the error per sample.

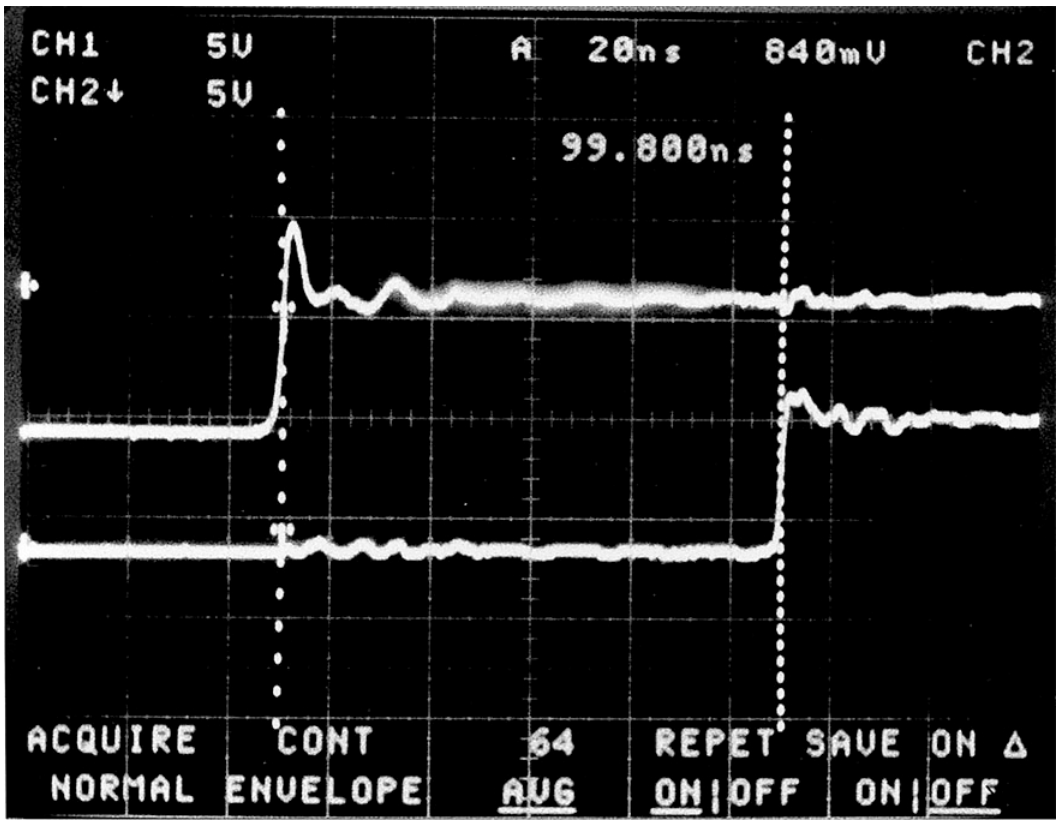

Figure 10: Oscilloscope photograph of the transient recorder trigger input and delay line test output. Conditions were: 7 Volt supply and delay line bias. Top trace shows trigger, bottom trace shows test output. Latency of trigger to output is 99.8 ns, corresponding to less than 200 ps per sample.

Delay line speed measurements were repeated over a range of delay line bias conditions in order to plot the speed dependency on bias. SPICE simulations were also performed under the same conditions. The results are plotted in figure 11. It is clear that SPICE modeled the general shape of the response quite accurately, but predicted higher performance than was actually achieved.

The absolute maximum speed measured was found to be a function of the power supply voltage. For this test, the analog delay adjustment input was set equal to the power supply voltage. With a five Volt supply, the maximum sample rate just exceeded 4 GHz, and with a 6 Volt supply, a 5 GHz rate was achieved with a small margin. At 7 Volts, the speed exceeded 5 GHz to the point where a closed loop control system could maintain the 5 GHz rate without difficulty. The measured and simulated speed dependence on power supply voltage is plotted in figure 12.

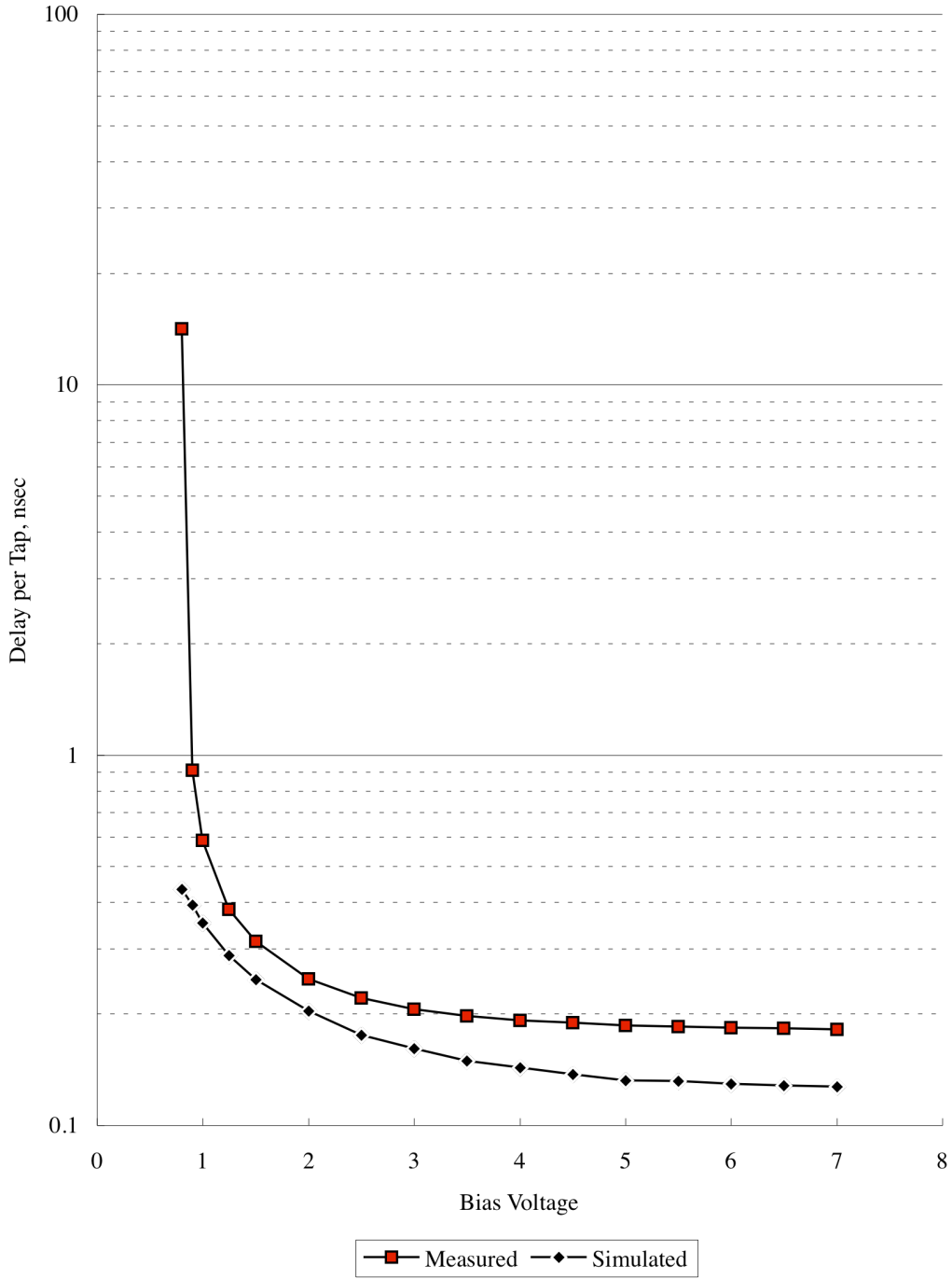

Figure 11: Delay Line Tap Latency vs. Bias Voltage.

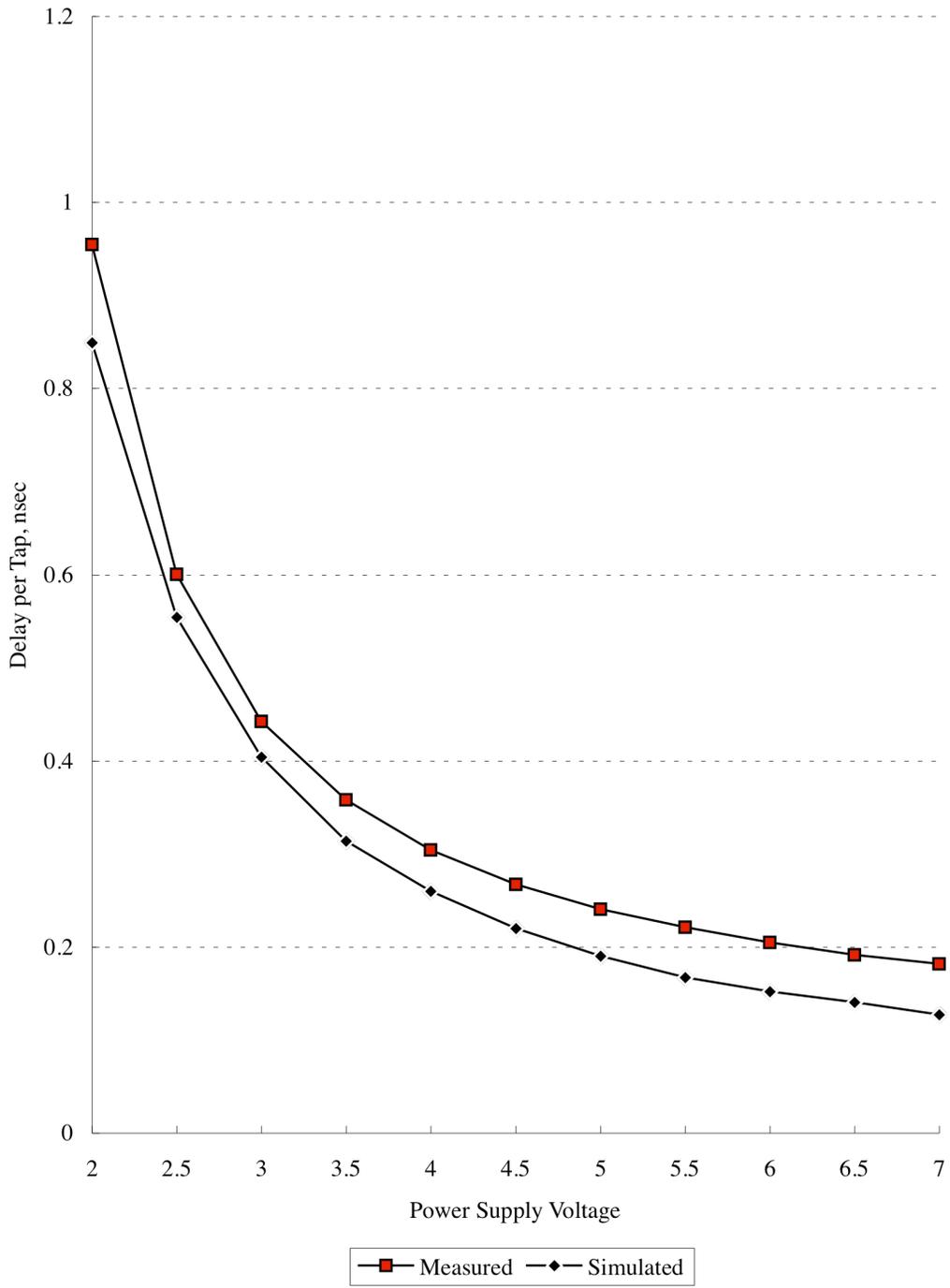

Figure 12: Delay Line Tap Latency vs. Power Supply Voltage.

*8.2. Internal sample clock width.*

An interesting measurement was made to determine the width of the internal sample and hold clocks. This measurement was made on a probe station (Allessi semi-automatic) using low capacitance "pico-probes," models 12C and 19C. These probes have a bandwidth of approximately 250 MHz, just below the 300 MHz of the Tektronix 4145 Oscilloscope used. A 600 MHz bandwidth Tektronix storage oscilloscope mainframe was borrowed briefly to support the notion that the Tek 4145 was not limiting the measurement.

This measurement was made with a 5 Volt power supply. Since there was no provision (i.e., probe pads) for this measurement, a careful cutting procedure was executed to expose a few of the 1.8 micron wide lines that carry the sample and hold clocks. Figure 13 shows an oscilloscope photo of one measured sample and hold clock signal. This measurement was taken in repetitive mode with no smoothing. The pulse is about 4 ns wide at 50% of maximum - very close to the SPICE simulated design value. The rise and fall times are about 1.6 ns. This is significantly slower than the ~200 ps predicted by SPICE, but is consistent with the 250 MHz bandwidth of the measurement apparatus. Notice that the peak of the waveform does not reach 5 Volts (the power supply used during this test) as the CMOS device should. This is due to the uncalibrated gain of the micro-probe.

*8.3. Input and output range.*

On the usual 6 volt supply, the measured effective operating limits of the analog input was 0 to 3.8 Volts. The limits on the output range were the same. These limits are simply due to the threshold of the n-channel sample and hold transistors given the substantial body effect they experience. For the n-well technology used, the body of the sample and hold transistors are always at ground potential, while their source follows the analog input. Would the body be connected to the analog input (possible if a p-well process were available), the signal range would presumably extend up to about 5.2 Volts (a $V_T$ at zero $V_{BS}$ below the positive supply), but with the side effect of increased input capacitance due to the added well to bulk capacitance.

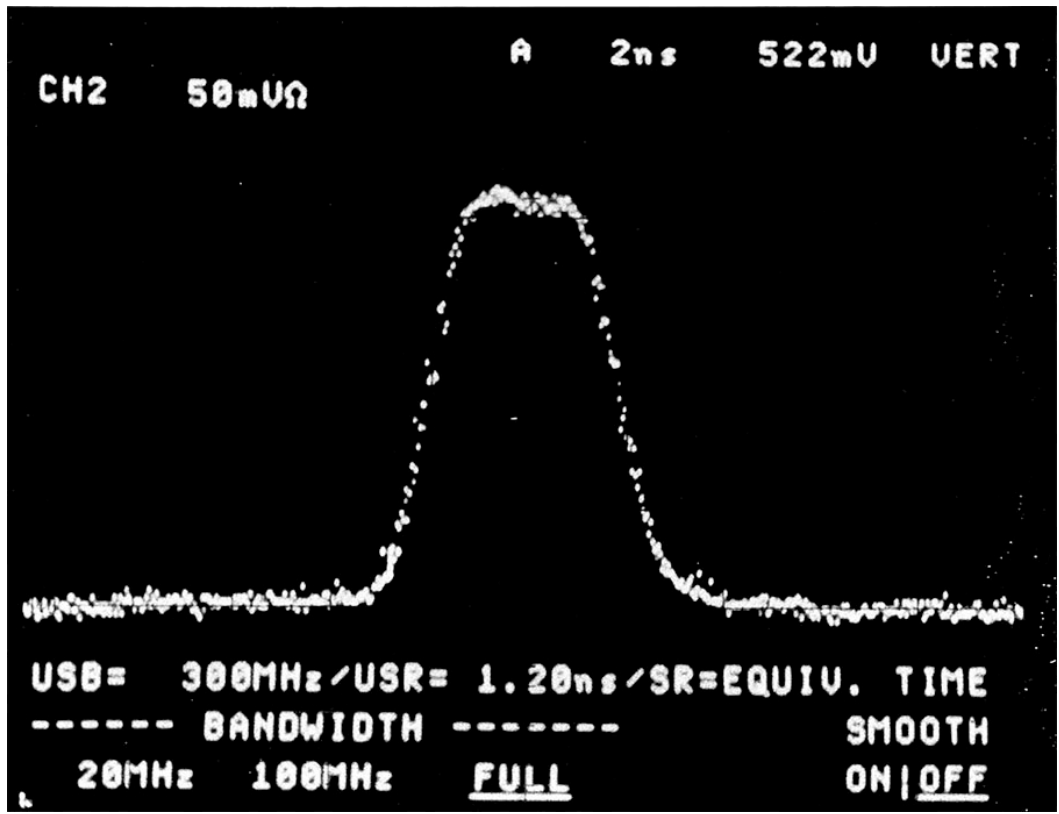

Figure 13: Oscilloscope photograph of an internal clock pulse probed from within the transient recorder circuit. Conditions were: 5 Volt supply and delay line bias. Clock width is seen to be approximately 4 ns. The ~1.6 ns clock edge rise and fall times that are observed are limited by the 300 MHz oscilloscope bandwidth.

### 8.4. Low frequency gain and linearity.

Low frequency gain was measured by inserting different D.C. voltage levels into the device and recording the output level. The measured low frequency gain of the device is nominally -1 (inverting). An oscilloscope is a poor device to measure large signal gain and linearity due to the limited D.C. coupled range at high gain. Better would be a computer automated measurement using a high resolution digitizer. A close measurement revealed slightly less than unity (inverting) gain near the extremes of range, perhaps by a few percent. At the extremes, the gain dropped to zero with extreme abruptness. At just a little away from the extremes, the gain was immeasurably different from unity. Figure 14 shows the results of this test.

### 8.5. Output noise.

A D.C. voltage from a low noise 50 Ohm source was applied such that the output signal excursions were minimized, permitting the oscilloscope to operate at maximum gain. The noise during the readout of representative samples was measured on the oscilloscope, and found to be about 1.2 mV rms. Conventional rms noise meters are unsuitable for this measurement due to the need to gate the measurement over small intervals of time. Other apparatus capable of making this measurement (i.e. "multi-channel analyzers") are difficult to use and of similar resolution. The very difficulty of the measurement, however, speaks of the success of the device, for 1.2 mV of noise over the 3.8 Volt range holds the promise of 3000:1 signal to rms noise. If accompanied by good cell to cell offset and timing uniformity, this would provide a 11.5 bits rms of dynamic range in a 5 GHz single shot device.

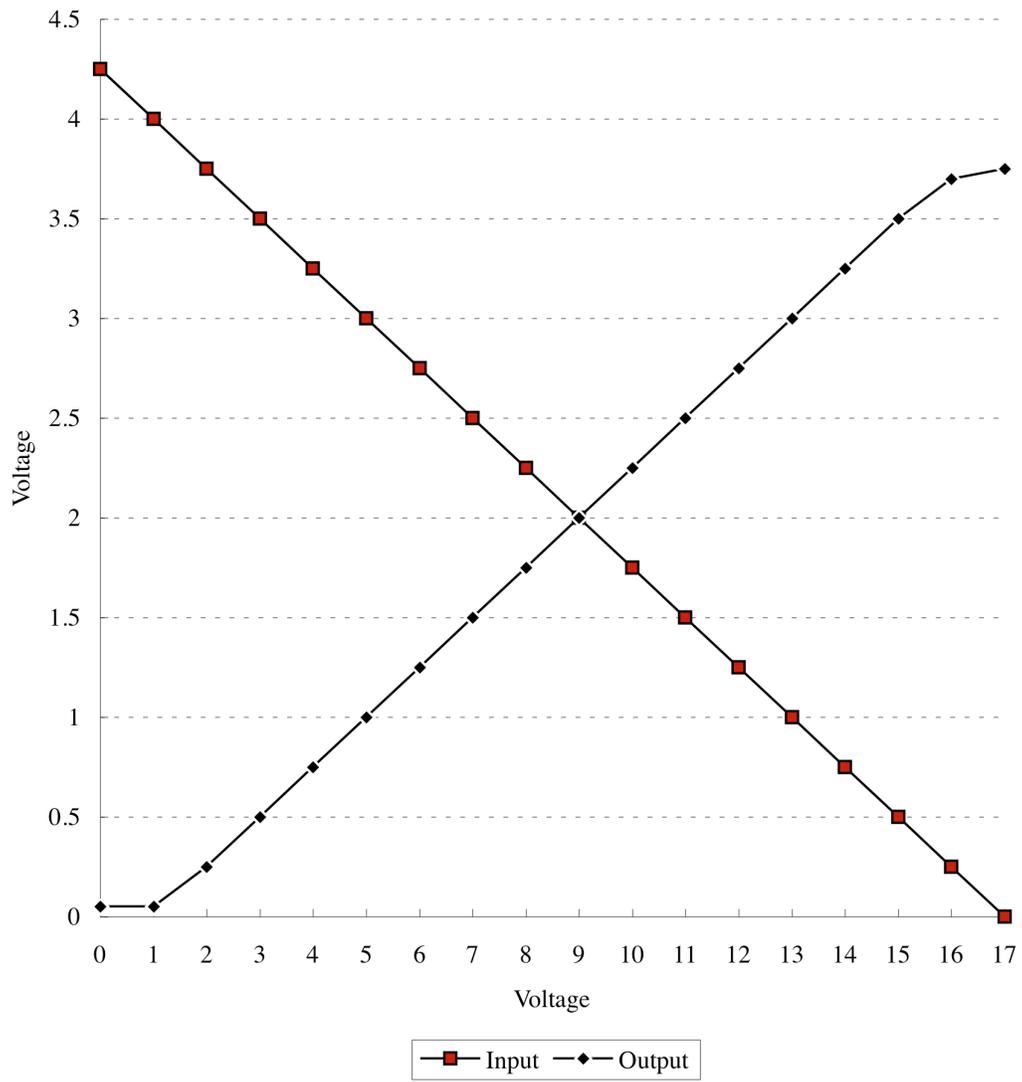

Figure 14: Transient Recorder Transfer Function.

*8.6. Baseline uniformity.*

As the device is made up of a multitude of individual sample and hold devices, their uniformity is an important issue. Uniformity has several modalities. Here, we are concerned with the so-called "spatial noise" that is independent of random noise. Measurements are carried out by averaging many acquisitions to remove the random noise components. What remains are systematic variations in the uniformity of capacitor values, charge injection from the sample and hold switches, and any other systematic effects.

When observing the baseline from all 512 samples, a shallow curvature was noted. The curvature dominated all random non-uniformity. The root mean square magnitude of this curvature is about 1.7 mV. The source of this curvature is unclear. One hypothesis was that it reveals decay over time of the stored values. This is because the readout rate is one or two hundred KHz, and the first cells are read out immediately, but the last cells are read out much later. This hypothesis was shown to be false, as varying the readout speed had little effect on the shape or magnitude of the curvature. The best hypothesis at present is that it is related to a processing dependent uniformity issue. The curvature could conceivably be monitored and subtracted from subsequent acquisitions, canceling its influence to a large degree.

*8.7. Input capacitance.*

Input capacitance was measured by injecting a fast edge through a 750 Ohm resistor into the test device. The resulting slow recorded edge is dominated by the resistance of the signal source impedance, the 750 Ohm test resistor, and the capacitance of the test chip. A measurement at the chip input using the oscilloscope would add the 12.5 pF scope probe capacitance to the time constant, so the sampler chip itself was used to determine the rise time. The result corresponded to an input capacitance of 15.7 pF, not including the board related parasitic. This amount of capacitance is too large to allow easy bandwidth measurements in excess of a few hundred MHz. In fact, the "700 ps" rise time signal source used (the best available) barely managed 2 ns rise time when loaded by the test chip.

*8.8. Analog bandwidth.*

After sample rate, the analog bandwidth is perhaps the next important performance figure. A compromise between input capacitance, noise and uniformity verses bandwidth is unavoidable in the design of a MOS sample and hold. By SPICE simulations and calculations, a peak bandwidth of about 1.8 GHz should be achieved by the test device when operated with low input voltage. In actual measurements, a performance of about 350 MHz was achieved. The measurement was conducted by injecting a pulse and observing the reconstructed rise and fall time.

A means of lowering the effective output impedance of the available signal source was necessary to achieve its limits in rise time. Thus, a 10 to 1 resistive divider, with an effective output impedance of about 5 ohms, was constructed for this test. This, unfortunately, also reduces the amplitude of the applied signal. Even when going to this extreme, the best pulse source available could still not manage better than about 1 ns rise time and, in fact, limited the measurement. Figure 15 shows one edge of a reconstructed pulse using the divider. There are about 5 samples measured from 10% to 90%. Given 200 ps per sample, this results in a rise time of 1 ns, for an approximate bandwidth of 350 MHz.

*8.9. Aperture uncertainty.*

Timing stability can be determined by observing the jitter in the delay line test output. The variation in total delay time is directly related to the stability of the delay line tap outputs. The oscilloscope was used to measured the total delay while set to an infinite persistence envelope mode. The width of the delay line test output rise time envelope, at the 50% amplitude point, gives the worst case peak to peak timing instability. Dividing by 6 yields a conservative rms jitter. The oscilloscope time base accuracy and noise in the oscilloscope trigger set a lower bound on the resolution of this measurement of about 0.5 ns peak to peak. Therefore, the contribution of the oscilloscope needs to be subtracted in quadrature from the total jitter. The relationship between the number of taps, the total rms jitter presuming 512 taps, and the rms jitter per tap is found to be:

$$\Delta t_{512} = 16 \Delta t_1 \sqrt{2}$$

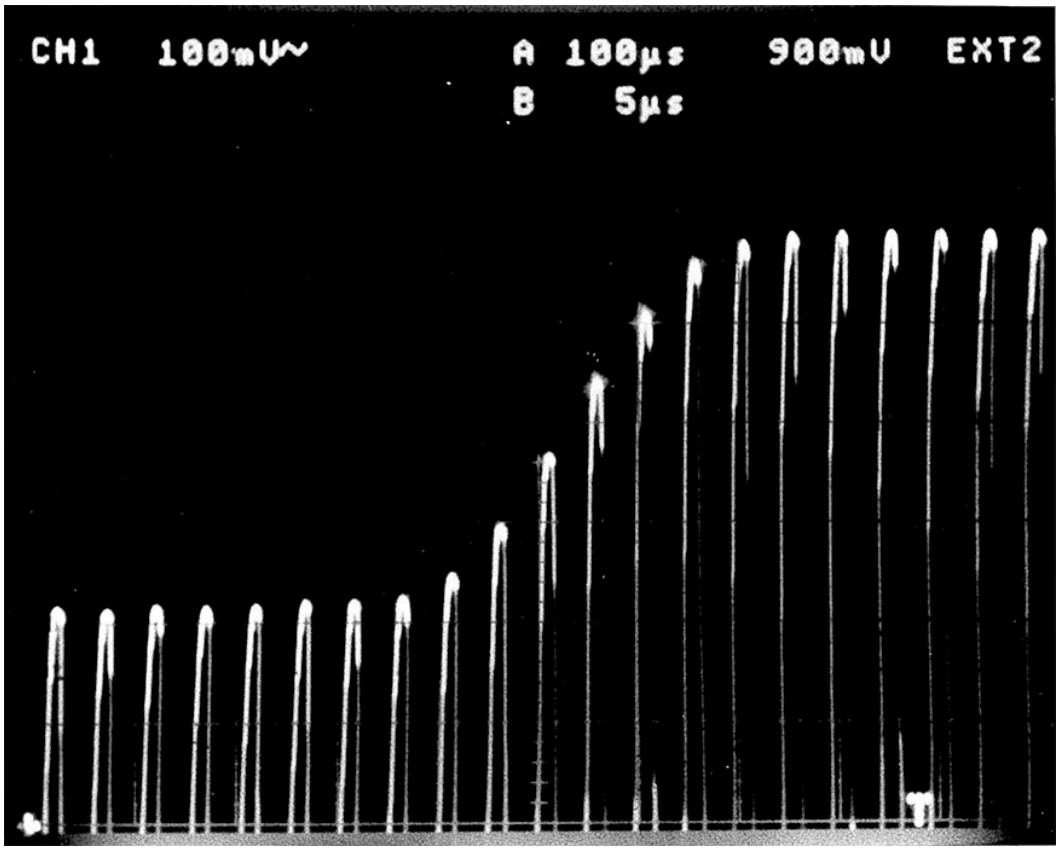

Figure 15: Oscilloscope photograph of the transient recorder analog readout of an acquired pulse edge. The recorder was set to 5 GHz acquisition rate. The reconstructed pulse edge shows a rise time (10% to 90%) of approximately 5 samples, corresponding to a 1 ns rise time.

On a 6 Volt supply, and with the delay control set to 0.9 Volts, the total delay was found to be 549 ns, for a sample frequency of about 0.962 GHz (1.04 ns period). The peak to peak envelope in the timing of the delay line test output, relative to the trigger input, was measured to be about 0.76 ns (see figure 16). If we take the worst case presumption that the oscilloscope contributes nothing to this measurement, then the rms jitter per sample is computed to be 5.6 ps. This amounts to 0.54 percent error on a single sample. If we subtract in quadrature the contribution by the oscilloscope, we arrive at an rms jitter for 512 samples of 95 ps and 4.2 ps on a single sample. This amounts to a 0.4 percent rms error over a single sample and 0.017 percent rms error over all 512 samples.

A second measurement was made with a much slower speed setting. On the same 6 Volt supply, the tap delay adjustment was set to 0.8 Volts, just above the transistor threshold. In this case, the total delay was found to be 6.58 microseconds, for a sample frequency of about 80 MHz, or a sample period of 12.5 ns. Using the same measurement technique, the timing non-uniformity was measured to be about 1.33 ns rms for 528 samples (see figure 17), or about 59 ps rms per tap . In this case, the oscilloscope contribution is negligible. This time, we find about 0.47 percent rms error for a single sample, and 0.02 percent rms over all 512 samples - almost unchanged from the 1 GHz sample rate case.

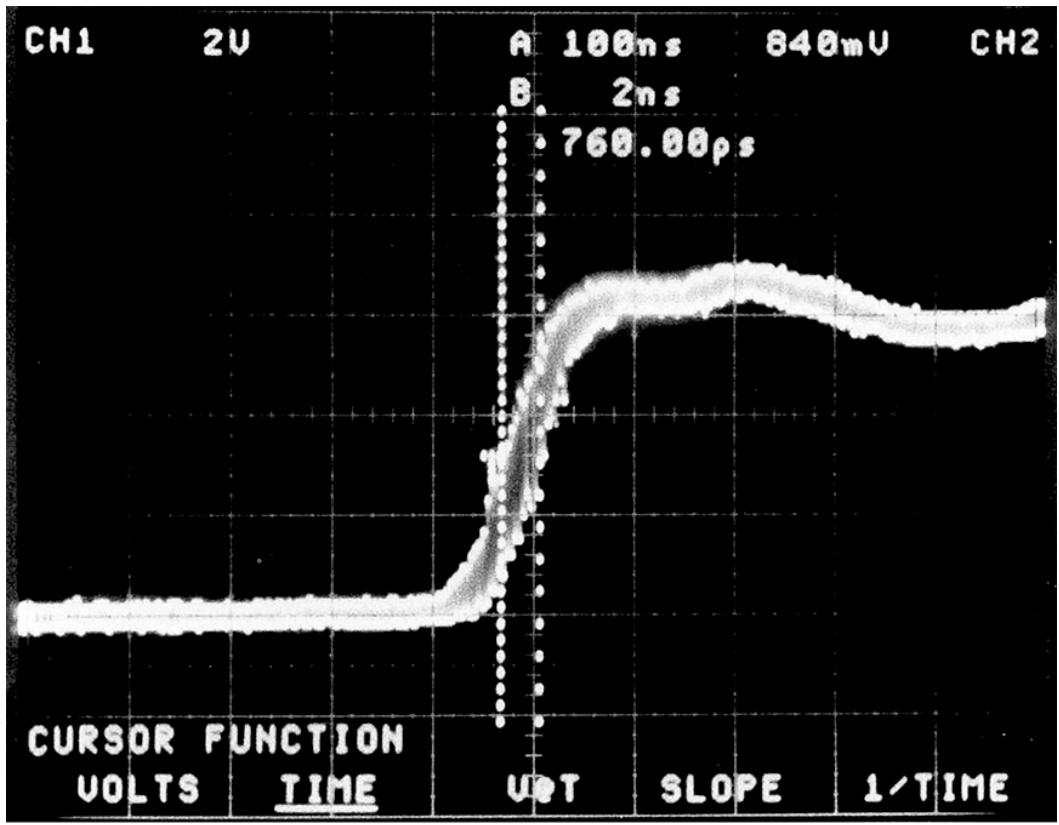

Figure 16: Envelope-mode oscilloscope photographs of the transient recorder delay line test output jitter, referenced to the trigger input. Conditions were: 6 Volt supply, 0.9 Volt delay line bias, 0.962 GHz sample rate. Peak to peak jitter is seen to be 760 ps, corresponding to an rms aperture uncertainty of less than or equal to 5.6 ps for one sample.

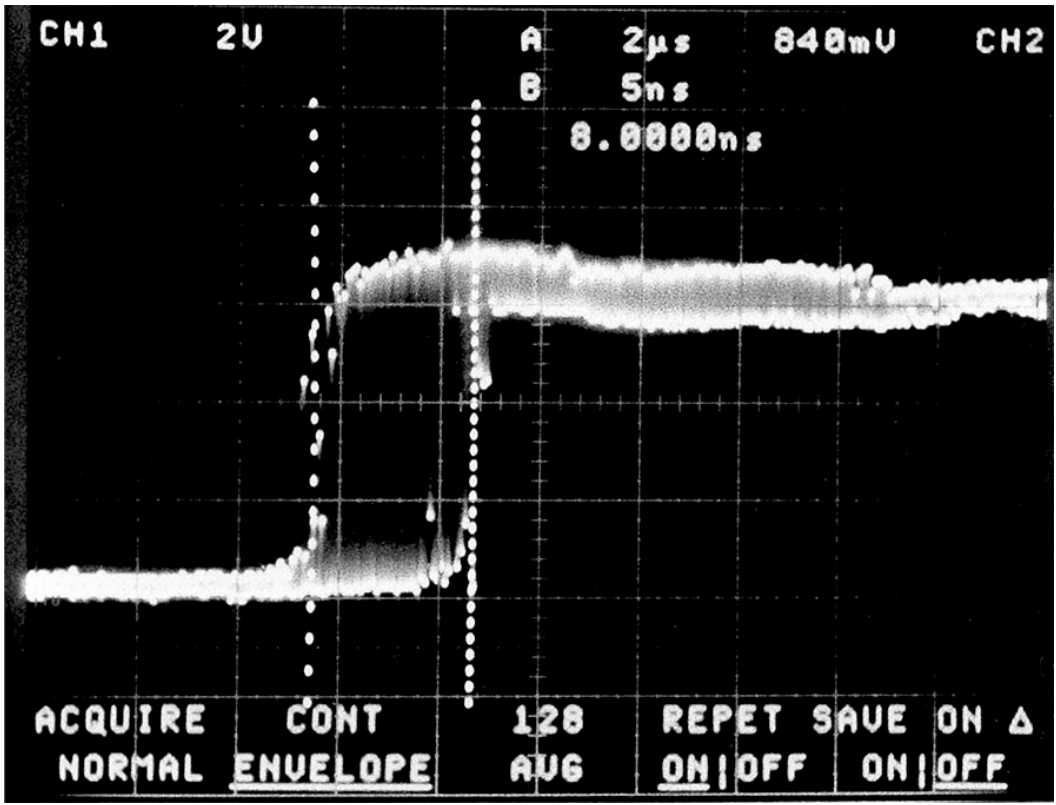

Figure 17: Envelope-mode oscilloscope photographs of the transient recorder delay line test output jitter, referenced to the trigger input. Conditions were: 6 Volt supply, 0.8 Volt delay line bias, 80 MHz sample rate. Peak to peak jitter is seen to be 8 ns, corresponding to an rms aperture uncertainty of less than or equal to 59 ps for one sample.

Using the above numbers, and extrapolating to the 5 GHz case, a jitter of about 0.7 ps rms per sample may be predicted. This would produce a peak to peak jitter at the delay line test output of about 0.095 ns - much lower than the 0.5 ns limit on the measuring instrument. The formula relating bandwidth to timing jitter and bits of vertical resolution is:

$$f = \left( 2^n \cdot \pi \cdot \Delta t \cdot \sqrt{6} \right)^{-1}$$

Where,

$f =$ Full scale sinewave frequency

$n =$ Bits of vertical resolution

$\Delta t =$ Timing jitter

At the maximum speed of 5 GHz, 6 bits of maximum effective resolution is predicted by this equation at a full-scale sine wave frequency of 2.9 GHz. The bandwidth falls as the number of bits of resolution demanded increases, dropping to 725 MHz at 8 bits, 181 MHz at 10 bits and 45 MHz at 12 bits. It is recognized that this limitation in bandwidth is diminished if the amplitude of the sine wave is reduced. This is because the maximum slope of the sine wave will drop. As a consequence, one common technique to increase effective bandwidth is to attenuate the input signal.

It is interesting to speculate about the source of timing uncertainty in this system. What is the source of the timing jitter and why does the percent rms jitter remain approximately equal over a wide range of tap delays? Conventional inverters have a threshold at which they switch that is considered to be at the crossing point of their D.C. input and output transfer functions. In an A.C. condition, there is noise associated with this threshold that arises from many sources. As the speed of the delay line is slowed, the edges associated with each delay tap become slower too - approximately linearly. The edge crosses through the threshold slower, and is susceptible to noise for a longer period of time. However, the noise on the threshold depends little on edge speed. Thus, we might expect the absolute amount of jitter to increase as the delay line is slowed, but might not expect the percent error to increase. This insight may permit us to design a system which is more immune to this increase in jitter. For example, if we believed that much of the jitter causing noise is due to noise on the power supply, a fully differential configuration may result in an improvement.

The performance characteristics of a ring shaped configuration running in common-stop trigger mode is also interesting to consider. As the ring oscillates, would one expect an accumulation of jitter to increase indefinitely? For example, the second time around the loop, the total jitter would increase by root 2, the fourth time around the loop, by a factor of 2, etc. But this is jitter with respect to the initiation of oscillation. In fact, in the common-stop configuration, the important point of time is the time of arrival of the stop trigger. Then the situation is the exact reverse of the 512 long open ended string - just that the well known point of time is the last sample, not the first.

Finally, does the observed jitter characteristics mean that larger strings or rings result in poorer timing? While the total error increases as you measure across more samples, the total error increases less than linearly, and so the ratio of total error over total time actually decreases. Therefore, large asynchronous sampling systems can remain useful.

Table 2. Performance Summary of the Fast Transient Recorder I.C.

| | |
|---|---|
| Technology: | 1.2 um CMOS |
| Die size: | 2.1 x 6.5 cm |
| Number of parallel channels per chip: | 4 |
| Number of sample and holds per channel: | 512 |
| Acquisition mode: | Asynchronous |
| Readout mode: | Sequential |
| Supply voltage: | 5-7 V |
| Power consumption per channel: | 10 mW |
| Maximum acquisition speed: | 5 GHz |
| Minimum acquisition speed: | 50 MHz |
| Readout speed: | 200 KHz |
| Input capacitance: | 15.7 pF |
| Acquisition bandwidth: | ~ 350 MHz |
| Output noise level (single sample): | 1.2 mV rms. |
| Cell to Cell variation (noise averaged out): | 1.7 mV rms. |
| Input range with 6 V supply: | 0 V - 3.8 V |
| Output range with 6 V supply: | 0.2 V - 3.8 V |
| Dynamic range, uncorrected: | ~ 11 bits rms. |
| Aperture uncertainty at 80 MHz, single sample: | 59 ps rms. |
| Percent uncertainty at 80 MHz, single sample: | 0.47 % rms. |
| Percent uncertainty at 80 MHz, 512 samples: | 0.02 % rms. |
| Aperture uncertainty at 1 GHz, single sample: | 4.2 ps rms. |
| Percent uncertainty at 1 GHz, single sample: | 0.4 % rms. |
| Percent uncertainty at 1 GHz, 512 samples: | 0.017 % rms |
| Aperture uncertainty at 5 GHz (extrapolated): | 0.7 ps rms. |
| Percent uncertainty at 5 GHz, single sample: | 0.35 % rms. |
| Percent uncertainty at 5 GHz, 512 samples: | 0.015 % rms. |

## 9. Possible Improvements and Further Experiments.

A common-stop architecture should be attempted in order to reveal the problems inherent in such a design. A fully differential delay line system would almost certainly be mandatory due to the need for an odd number of inversions in the loop to cause oscillation, but an even number of inversions per tap to maintain tap to tap uniformity. It is expected that a common-stop architecture would suffer in density due to the substantial overhead associated with recording the position of the write clock edge. It could also be expected that small discontinuities in the baseline would be detectable due to the presence in the layout of corners in the loop. Process gradients may also cause triangular shaped non-uniformity in the baseline. Creating a layout with a uniform circle or oval arrangement of sample and hold elements may eliminate the corner effects, but would severely limit density. A closed loop sample rate control to lock the frequency of oscillation to some slow external reference would be necessary in a practical system. This could be accomplished with the techniques developed for phase locked loop systems.

Attaining higher sample rates would require the use of a superior technology. Few manufacturers offer sub-micron processes that include a high density capacitor layer. However, it may be possible to use a process lacking "discrete" capacitors provided that the circuit can tolerate the use of small parasitic capacitors or FET capacitors. Triple layers of metal might be necessary to make up for the lack of a second polysilicon layer. Dynamic range and linearity would likely suffer, but there would be some applications that would gladly make this tradeoff. The circuit may also benefit from the use of a BiCMOS process. The use of BiCMOS stages may permit higher speeds and faster sample and hold clock edges.

Analog bandwidth would also benefit from a superior technology. If a fully differential sample and hold technique were used, the size of the storage capacitors could likely be reduced while maintaining good resolution. By reducing the storage capacitor size, the RC time constant is reduced. By combining smaller capacitors with smaller, shorter channel switches, the parasitic associated with the sum of the switch source diffusions can be reduced. The input capacitance will drop and the bandwidth will increase and . The use of large diameter bond wires or ribbon bonding, and aggressive high frequency board or hybrid design would permit higher system bandwidths to be reached. Finally, since the individual voltage sampling track and holds are predicted (by SPICE) to be much faster than the measured performance of the over-all system, it may be that a current mode input followed by internal current to voltage conversion could also result in an increase in system bandwidth.

## 10. Conclusions.

This research has experimentally demonstrated that an analog transient recorder with high resolution and 5 GHz sample rates can be achieved using a modest CMOS process. The transient recorder uses a switched-capacitor method to store the input transient onto 512 capacitors. The high sample rate was achieved by the development of a novel asynchronous delay line write clock generation system. Using 16 way interleaving, 512 clocks of as little as 3.2 ns width and 200 ps apart were generated. Aperture uncertainty at the 5 GHz sample rate was extrapolated from measurements and estimated to be 0.7 ps rms. Measured bandwidth was limited to 350 MHz, though some indications of higher achievable bandwidth were noted. Signal to noise was found to be 2000:1. The 2.1 by 6.5 mm die contains 4 parallel channels and uses only 40 mW of power at full speed. The promising results that this first experimental device produced bodes well for future research.

# 11. References.